\documentclass[twocolumn,english,aps,prb,floatfix,preprintnumbers,showpacs,amsfonts,amssymb]{revtex4}
\usepackage[T1]{fontenc}
\usepackage[latin1]{inputenc}
\usepackage{amsmath}
\usepackage{amssymb}

\makeatletter

\providecommand{\LyX}{L\kern-.1667em\lower.25em\hbox{Y}\kern-.125emX\@}

\usepackage{graphicx}
\usepackage{graphicx}
\usepackage{graphicx}
\usepackage{amscd}
\usepackage{bm}

\usepackage{babel}
\makeatother
\begin{document}

\title{Optical conductivity of charge carriers interacting with a two-level
systems reservoir}

\author{A. Villares Ferrer}

\affiliation{Instituto de F\'{\i}sica, Universidade Federal Fluminense, Av. Gral
Milton Tavares de Souza s/n, Niter\'{o}i 24210-346, Rio de Janeiro,
Brazil.}

\author{A.O. Caldeira}

\affiliation{Instituto de F\'{\i}sica {}``Gleb Wataghin'', Departamento de
F\'{\i}sica da Mat\'{e}ria Condensada, Universidade Estadual de
Campinas, CP 6165, Campinas 13083-970, S\~{a}o Paulo, Brazil.}

\author{C. Morais Smith}

\affiliation{Institute for Theoretical Physics, University of Utrecht, Leuvenlaan
4, 3584 CE, Utrecht, The Netherlands.}

\date{\today{}}

\begin{abstract}
Using the functional-integral method we investigate the effective
dynamics of a charged particle coupled to a set of two-level systems
as a function of temperature and external electric field. The optical
conductivity and the direct current (dc) resistivity induced by the
reservoir are computed. Three different regimes are found depending
on the two-level system spectral function, which may lead to a non-Drude
optical conductivity in a certain range of parameters. Our results
contrast to the behavior found when considering the usual bath of
harmonic oscillators which we are able to recover in the limit of
very low temperatures. 
\end{abstract}


\maketitle

\section{Introduction}

The possibility of manipulating physical systems at length scales
where quantum effects become important has attracted much attention
over the last years. This scenario has stimulated considerable effort
in the study of quantum open systems, firstly because of its relevance
to the phenomenon of quantum coherence and secondly because it might
not be entirely obvious the way in which we can obtain information
about the quantum dynamics of the system of interest coupled to its
environment. One of the strategies broadly used in solving this problem
consists in the replacement of the environment by an approximate model.
This must be done in such a way that after tracing the environment
coordinates out, the problem can be formulated only in terms of the
variables of the system of interest.\cite{FeyVer} It is remarkable
that most of the environments can be represented either in terms of
a bath of oscillators\cite{AOphysa,AJLrmodphy}, when the physics
is dominated by the delocalized modes, or by an spin bath\cite{PSrepprphy},
when the localized modes play the major role. In this paper we will
be interested in another type of thermal bath that can be thought
of as the projection of the usual oscillator modes onto their two
lowest lying levels. At very low temperatures, these truncated two-level
systems (TLSs) have the same properties as the usual harmonic oscillators,
that is, the two baths exhibit the same quantum limit. However, as
we will show, they strongly differ in the classical regime, at high
temperatures.

It is now well known that when the oscillator model with linear spectral
density is used to mimic a thermal bath interacting with a quantum
particle, the wave packet associated with the latter undergoes a damped
motion, exactly as in the classical problem.\cite{AOphysa} In this
situation, and within the long time approximation, the average over
the environment variables results in an equation of motion for the
particle without memory effects. Therefore, it directly follows that
the transport properties of the quantum particle can be simply described
in terms of the damping and diffusion coefficients. As a consequence,
the optical conductivity $\sigma (\omega )$ of a single particle
coupled to an oscillator bath has, in the so-called ohmic case, only
an incoherent part, which has a Drude-like form. In this case the
Lorentzian width is determined by a \textit{temperature independent}
damping constant.

The phenomenological approach of representing the environment by an
oscillator bath was successfully used in the study of dissipative
effects in macroscopic quantum coherence \cite{AJLrmodphy} (the spin-boson
model) and macroscopic quantum tunneling \cite{AOannal}. Moreover,
there are particular situations, see {[}\onlinecite{ferenvi}{]} for
instance, in which the oscillator model can be derived from microscopic
theories just following the prescriptions of Feynman and Vernon \cite{FeyVer}.
That is also the case of solitons, whose transport properties can
be investigated using the collective coordinate quantization scheme
\cite{soli,antife,skyr}. In those cases the effective equation of
motion for the center of mass of the soliton leads, in the long time
regime, to a \textit{temperature dependent} damping constant. The
form of the damping constant is such that the optical conductivity
of a system of non-interacting solitons has again a Drude-like form,
and in the low temperature limit correctly reproduces the finite free
particle Drude weight at zero frequency. This behavior is completely
general for solitons and therefore independent of the non-linear field
theory that supports these localized solutions. From {[}\onlinecite{AOphysa}{]}
and {[}\onlinecite{soli}-\onlinecite{antife}{]} we can conclude
that, within the long time regime, the optical properties of quasi-particles
coupled to an oscillator bath have always the trivial Drude form.
Therefore, if results at variance with the latter are obtained in
$\sigma (\omega )$ measurements, they can not be attributed solely
to the above-mentioned particle-reservoir interaction. In those cases
more complicated models are required\cite{imadather} or alternative
thermal bath descriptions should be employed.\cite{PSrepprphy}

In this paper, our main goal is to present a different kind of thermal
reservoir which, in addition to a dissipative dynamics for the charge
carriers, induces a non-trivial optical conductivity. Our starting
point will be the problem of a single particle subject to a complex
potential which can be represented as a distribution of local two-level
systems. Using the well known Feynman-Vernon formalism\cite{FeyVer}
we are able to traced out the environment modes and obtain the reduced
density operator of the particles. This operator will be expressed
in terms of an effective action containing all the information about
the coupling of the carriers to the reservoir and provides us with
an equation of motion for the particles from which the transport properties
can be computed. In this way it will be shown that the optical conductivity
of this system has a temperature dependent non-Drude behavior with
a very rich structure which contrasts with the oscillator bath transport
properties.

This paper will be divided as follows. In Sec.\ II we present a model
describing a single particle interacting with a set of TLSs. In Sec.\ III
an effective equation of motion describing the particle dynamics will
be derived and in Sec.\ IV the optical conductivity of the system
is calculated considering different characteristics of the thermal
reservoir. Finally, in Sec.\ V, we discuss our results and present
the conclusions.

\section{The Model}

As it was already mentioned, we will be interested in the transport
properties of a single particle coupled to a generic TLSs reservoir.
This problem can be completely described by the hamiltonian\begin{equation}
H=H_{0}+H_{r}+H_{i},\label{eq:HTot}\end{equation}
 where $H_{0}$ stands for a particle placed in an external electric
field, and it is given by \begin{equation}
H_{0}=\frac{{\hat{p}}^{2}}{2M}+exE.\label{eq:hamiliv}\end{equation}
 The distribution of TLSs playing the role of thermal reservoir in
(\ref{eq:HTot}) is denoted by $H_{r}$ and may be expressed in the
$k-$space as \begin{equation}
H_{r}=\sum _{k=1}^{N}\frac{{\hbar \omega _{k}}}{2}\sigma _{zk},\label{eq:hamires}\end{equation}
 where $\sigma _{zk}$ is the z-Pauli matrix. Finally, the interaction
between the particle of interest and the thermal bath $H_{i}$ will
be taken in such a way to induce transitions between the states of
each TLS. A suitable choice is \begin{equation}
H_{i}=-x\sum _{k=1}^{N}J_{k}\sigma _{xk},\label{eq:hamiint}\end{equation}
 where $J_{k}$ and $\sigma _{xk}$ are, respectively, the coupling
constant and the x-Pauli matrix.

\begin{figure}
\begin{center}\includegraphics[  clip,
  scale=0.43]{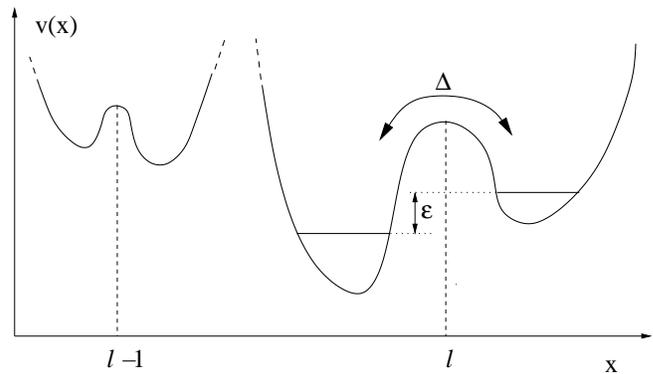}\end{center}

\caption{\label{potpot}Complex potential with multiple local quartic structure.
The effective TLS at site $l$ is described in terms of the detuning
$\varepsilon $ between the wells ground states and a typical matrix
element for the tunneling processes $\Delta $.}
\end{figure}

Although the problem defined by Eqs.\
(\ref{eq:HTot})-(\ref{eq:hamiint}) was not derived from a microscopic
description of a concrete physical system, it is still very useful
because there are many systems which, under certain circumstances,
behave as a truncated TLS. Indeed, in the process of modeling defects
in crystalline solids or amorphous materials, one has to deal with
a distribution of locally quadratic-plus-quartic potentials, as it
is shown in Fig.\
\ref{potpot}. In fact, if each local double-well in the distribution
has fairly separated minima, and the typical energy scale $\hbar \omega _{c}$
(obtained by the harmonic approximation about each minimum and assuming
they do not differ much) is such that $\hbar \omega _{c}>>kT$ (see
Refs.\ {} {[}\onlinecite{AJLrmodphy}{]} and {[}\onlinecite{Truncation}{]}
for details) all the locally quartic potentials at positions $l$
can be effectively described as two-level systems and \begin{equation}
H_{l}=-\frac{{1}}{2}\left(\Delta _{l}\sigma _{xl}-\varepsilon _{l}\sigma _{zl}\right),\label{eq:lham1}\end{equation}
 where $\Delta _{l}$ is a typical matrix element for the tunneling
process between the two minima and $\varepsilon _{l}$ is the {}``detuning''
between the ground states in the two wells. This fact suggests a possible
phenomenological description for a particle moving in a complex potential
as shown in Fig.\ 1, where the source of dissipation is the induced
transition between the states of the TLSs ensemble and may have the
form (\ref{eq:hamiint}). However, in the course of this procedure
one should keep in mind that although the rate $\varepsilon _{l}/kT$
may have any value, the TLSs representation is valid only if $\hbar \omega _{c}>>kT\sim (\hbar \Delta _{l},\varepsilon _{l})$.
Therefore the high temperature limit should be understood as $\hbar \omega _{c}>>kT>\hbar \Omega $
where $\Omega $ is the cutoff frequency of the system, which can
be assumed to be of the order of max$(\hbar \Delta _{l},\varepsilon _{l})$.

Knowing the details of the model describing the TLSs array we can
go further and study the dynamics of a particle coupled to this environment.

\section{Effective Particle dynamics}

\subsection{The Influence Functional}

This section is devoted to investigating the effective dynamics of
a particle interacting with the TLS thermal reservoir. We will use
the Feynman-Vernon functional integral approach, which begins with
the calculation of the reduced density operator of the particle of
interest, namely \begin{equation}
\rho (x,y,t)=Tr[\langle x|e^{-i\frac{{Ht}}{\hbar }}\rho (0)e^{i\frac{{Ht}}{\hbar }}|y\rangle ],\label{eq:res1}\end{equation}
 where we are using the coordinate representation for the particle
states, $Tr$ denotes the trace over the bath modes and $H$ is the
Hamiltonian of the total system given by (\ref{eq:HTot}).

The density operator of the whole system at time $t=0$ will be assumed
to be separable, $\rho (0)=\rho _{p}(0)\rho _{r}(0)$, where $p$
and $r$ denote the particle and the reservoir, respectively. The
reduced density operator (\ref{eq:res1}) can be written as \[
\rho (x,y,t)=\int dx'\int dy'\rho _{p}(x',y',0){\mathcal{J}}(x,y,t;x'y',0),\]
 where the superpropagator ${\mathcal{J}}$ has the form \begin{equation}
{\mathcal{J}}=\int _{x'}^{x}{\mathcal{D}}x(t')\int _{y'}^{y}{\mathcal{D}}y(t')e^{\frac{{i}}{\hbar }(S_{0}[x]-S_{0}[y])}{\mathcal{F}}[x,y].\label{eq:sup2}\end{equation}
 In the expression above \[
S_{0}[z]=\int _{0}^{t}\left[\frac{M}{2}\dot{z}^{2}(t')+ez(t')E(t')\right]dt'\]
 corresponds to the action of a free particle in the presence of an
electric field and ${\mathcal{F}}$ denotes the influence functional,
which is given by \begin{eqnarray}
{\mathcal{F}}[x,y] & = & Tr\left[\rho _{r}(0)\: \left(T\, \exp \frac{{i}}{\hbar }\int _{t_{0}}^{t}\widetilde{H}_{i}(y(t'))dt'\right)\right.\nonumber \\
 &  & \nonumber \\
 &  & \left.\left(T\, \exp -\frac{{i}}{\hbar }\int _{t_{0}}^{t}\widetilde{H}_{i}(x(t'))dt'\right)\right],\label{eq:influ1}
\end{eqnarray}
 where $T$ indicates time-ordered product and \[
\widetilde{H}_{i}(z)=e^{^{\frac{{i}}{\hbar }H_{r}t}}H_{i}(z(t'))e^{^{-\frac{{i}}{\hbar }H_{r}t}}.\]

The central quantity in this method is the influence functional defined
in Eq.\ (\ref{eq:sup2}). Our next step is to derive an explicit
expression for it. Although this quantity has already been evaluated
through different approaches for interactions of the form (\ref{eq:hamiint})\cite{AOtwlevel},
here we will sketch its derivation for the sake of completeness. 

In order to proceed, we will assume that the interaction strength
is weak enough, such that we may expand Eq.\ (\ref{eq:influ1}) in
powers of $J_{k}$ and retain only terms up to second order. We then
obtain \begin{widetext}\begin{eqnarray}
\mathcal{F}[x,y]=1-\frac{1}{\hbar ^{2}}\int _{0}^{t}dt'\int _{0}^{t'}dt''
&  & \left\{ \left\langle \widetilde{H}_{i}(x(t'))\widetilde{H}_{i}(x(t''))
\right\rangle +\left\langle \widetilde{H}_{i}(y(t''))\widetilde{H}_{i}(y(t'))
\right\rangle -\left\langle \widetilde{H}_{i}(y(t'))\widetilde{H}_{i}(x(t''))
\right\rangle \right.\nonumber \\
 &  & -\left.\left\langle \widetilde{H}_{i}(y(t''))\widetilde{H}_{i}(x(t'))
\right\rangle \right\} ,
\label{eq:infexp}
\end{eqnarray}
where the average value of an observable \(A\) is given by
\[
\left\langle A\right\rangle =\left[2^{N}{\displaystyle \prod _{k=1}^{N}}
\cosh \left(\frac{\hbar \omega _{k}}{2k_{B} T}\right)\right]^{-1}
Tr\left[e^{-H_{r}/k_{B} T}\, A\right].
\]

After tracing the reservoir degrees of freedom from Eq.\
(\ref{eq:infexp}), the influence functional acquires the
form,\begin{equation} \mathcal{F}[x,y]=1-\frac{1}{\hbar ^{2}}\int
_{0}^{t}dt'\int _{0}^{t'}dt'' {\displaystyle \sum
_{k=1}^{N}}J_{k}^{2}\left\{ f(x,y)\, \cos \left[\omega
_{k}(t'-t'')\right] -ig(x,y)\, \tanh \left(\frac{\hbar \omega
_{k}}{2k_{B} T}\right) \sin \left[\omega
_{k}(t'-t'')\right]\right\} ,\label{eq:ffexx}\end{equation} where
\[
f(x,y)=x(t')x(t'')+y(t'')y(t')
-y(t')x(t'')-y(t'')x(t'),
\]
and
\[
g(x,y)=x(t')x(t'')-y(t'')y(t')
+y(t')x(t'')-y(t'')x(t').
\]
Equation (\ref{eq:ffexx}) may be simplified by introducing a
set of coordinates corresponding to the particle center of mass \(q\)
and relative coordinate \(\xi \). Therefore, \(x=q+\xi /2\),
\(y=q-\xi/2\), and the influence functional reads
\begin{equation}
\mathcal{F}[x,y]=1-\frac{1}{\hbar ^{2}}\int _{0}^{t}dt'\int _{0}^{t'}dt''
{\displaystyle \sum _{k=1}^{N}}J_{k}^{2}\left\{ \xi(t')\xi(t'')\,
\cos \left[\omega _{k}(t'-t'')\right]-2i\, q(t')\xi(t'')
\tanh \left( \frac{\hbar \omega _{k}}{2k_{B} T} \right)
\sin \left[\omega _{k}(t'-t'')\right]\right\}. \label{eq:infvel1}
\end{equation}

In the majority of cases considered we do not have enough
information about the system which would allow us to perform the
above summation in \(k\). In order to overcome this difficulty,
one usually introduces a phenomenological spectral density
function, which correctly describes the bath in some well-known
limit.\cite{AOphysa} Because we intend to follow this strategy, we
will first calculate the form of the spectral density function
associated to the two-level system reservoir and  will replace it
by some phenomenological guess afterwards.

The spectral function can be obtained from
\begin{equation}
J(\omega )=Im\left(\mathfrak{F}\left\langle -i\Theta (t-t')
\left[F(t),F(t')\right]\right\rangle \right),
\label{eq:specdeng}
\end{equation}
where \(F(t)\) is the force produced by the particle over the
thermal bath , \(\Theta (t)\) is the usual step function, and
\(\mathfrak{F}\) stands for the Fourier transform. From Eq.\
(\ref{eq:hamiint}) it is straightforward to demonstrate that
\(F(t)=\sum _{k}J_{k}\sigma _{xk}\). Substituting then the
expression for the force into Eq.\ (\ref{eq:specdeng}) and using
Eq.\ (\ref{eq:hamires}) we obtain
\begin{equation}
J(\omega ,T)={\displaystyle \sum _{k=1}^{N}\pi
J_{k}^{2}\tanh \left(\frac{\hbar \omega _{k}}{2k_{B} T}\right)\,
\delta (\omega _{k}-\omega )}.
\label{eq:specden1}
\end{equation}

Notice that the above expression is completely different from the
oscillator spectral function. Here, all modes with energy well
below \(k_{B} T\) contribute less effectively to the spectral
density than the {}``unoccupied'' modes above the thermal energy.
Therefore, any attempt to replace (\ref{eq:specden1}) by a
phenomenological guess should preserve this property.

If we now replace Eq.\ (\ref{eq:specden1}) into Eq.\ (\ref{eq:infvel1}),
we obtain the influence functional
\begin{equation}
\mathcal{F}=\exp \left\{-\frac{1}{\hbar }\int _{0}^{t}dt'
\int _{0}^{t'}dt''\int _{0}^{\infty }d\omega \, \frac{{J(\omega ,T)}}{\pi}
\left\{ \coth \left(\frac{\hbar \omega }{2k_{B} T}\right)\xi(t')\xi(t'')
\cos \left[\omega (t'-t'')\right]-2iq(t')\xi(t'')
\sin \left[\omega (t'-t'')\right]\right\} \right\},\label{eq:infxx}
\end{equation}
\end{widetext} where we have re-exponentiated the second order expansion used in
(\ref{eq:infexp}).

Next, we substitute Eq.\ (\ref{eq:infxx}) into Eq.\ (\ref{eq:sup2})
to obtain the superpropagator for the particle of interest, namely
\begin{eqnarray*}
\mathcal{J} & = & \int {\mathcal{D}}\xi \int {\mathcal{D}}q\: \exp \left\{ \frac{{i}}{\hbar }S_{eff}[q,\xi ]\right\} \times \\
 &  & \exp \left\{ -\frac{{1}}{\hbar }\int _{0}^{t}\int _{0}^{t'}\Phi (t'-t'')\xi (t')\xi (t'')dt'dt''\right\} .
\end{eqnarray*}
 In the above equation, the effective action is given by \begin{eqnarray}
S_{eff} & = & \int _{0}^{t}dt'\left[M\dot{q}(t')\dot{\xi }(t')+e\xi (t')E(t')\right.\label{eq:effact}\\
 &  & +\int _{0}^{t'}\left.\Lambda _{1}(t'-t'')q(t')\xi (t'')dt''\right],\nonumber 
\end{eqnarray}
 where \[
\Lambda _{1}=\frac{2}{\pi }\int _{0}^{\infty }J(\omega ,T)\, \sin [\omega (t'-t'')]d\omega ,\]
 and \[
\Phi =\frac{1}{\pi }\int _{0}^{\infty }J(\omega ,T)\, \cos [\omega (t'-t'')]\coth \left(\frac{\hbar \omega }{2k_{B}T}\right)d\omega .\]

After having traced the environment coordinates, we can derive from
the effective action (\ref{eq:effact}) an equation of motion for
the time evolution of the particle center of mass $q$ and for the
width $\xi $ of the wave packet associated to it, \begin{equation}
M\ddot{q}(t)+\int _{0}^{t}\Lambda (t-t')\dot{q}(t')dt'=eE(t),\label{eq:cofmass}\end{equation}
\begin{equation}
M\ddot{\xi }(t)+\int _{0}^{t}\Lambda (t-t')\dot{\xi }(t')dt'=0,\label{eq:width}\end{equation}
 where we have performed an integration by parts in order to explicitly
show the viscous force of the TLSs reservoir acting on the particle
and\begin{equation}
\Lambda (t-t')=\frac{{2}}{\pi }\int _{0}^{\infty }\omega ^{-1}J(\omega ,T)\cos [\omega (t-t')]d\omega .\label{eq:Lamb2}\end{equation}

The last step consists in solving Eqs.\ (\ref{eq:cofmass}) and (\ref{eq:width}).
However, we must first specify the form of the spectral density. Because
we intend to keep the problem as general as possible, we will choose
a form for the spectral density which retains the functional $T$-dependence
given by Eq.\
(\ref{eq:specden1}). We then assume that \begin{equation}
J(\omega ,T)=\frac{\pi }{2}\, \overline{\gamma }\, \omega ^{s}\tanh \left(\frac{\hbar \omega }{2k_{B}T}\right)\; \Theta (\Omega -\omega ),\label{eq:spectral}\end{equation}
 where $\Omega $ is the cutoff frequency already introduced in the
discussion of the model, $\overline{\gamma }$ is a constant defining
the particle coupling strength to the TLSs and $s$ is a c-number.
The reader must be warned that $\overline{\gamma }$ used in this
paper is actually $4\gamma /\pi $ where $\gamma $ is the usual relaxation
constant of the particle motion.

Expression (\ref{eq:spectral}) retains the main properties of the
functional form (\ref{eq:specden1}), namely the fact that the temperature
determines which are the statistically relevant modes. Therefore,
for a given value of $\Omega $ and depending on the temperature,
the particle may simply have no interaction, on average, with the
reservoir. Notice that this choice is at variance with the one employed
in Ref. {[}\onlinecite{Nmakri}{]} whre the author maps the system
onto a bath of oscillators.

By replacing Eqs.\ (\ref{eq:Lamb2}) and (\ref{eq:spectral}) into
(\ref{eq:cofmass}), we obtain the following equation for the velocity
of the particle center of mass: \begin{equation}
\dot{v}(t)+\overline{\gamma }\int _{0}^{t}\Gamma (t-t')\, v(t')dt'=\frac{{eE(t)}}{M},\label{eq:velomed}\end{equation}
 with the damping function \begin{equation}
\Gamma (t-t')=\int _{0}^{\Omega }\omega ^{s-1}\, \tanh \left(\frac{{\hbar \omega }}{2k_{B}T}\right)\cos [\omega (t-t')]\, d\omega .\label{eq:gamma}\end{equation}

As it can be observed from Eq.\ (\ref{eq:velomed}), after tracing
the two-level system reservoir coordinates, we have obtained an equation
of motion for the particle center of mass in which the thermal bath
has the same effect as that of a \emph{viscous fluid.} It should be
noticed that for $s=1$ and zero temperature, Eq.\ (\ref{eq:gamma})
reduces to the oscillator-bath damping function, which is memoryless
in the limit $\Omega \rightarrow \infty $ or, in other words, in
the long time regime. Indeed, in this case \[
\Gamma (t-t')=\lim _{\Omega \rightarrow \infty }\int _{0}^{\Omega }\cos (\omega [t-t'])d\omega =\pi \delta (t-t'),\]
 indicating that the damping is a purely instantaneous function.

However, if $s\neq 1$ we realize that even at zero temperature and
with the cutoff frequency going to infinity, it is impossible to obtain
a damping function without memory. In this situation the problem becomes
non-Markovian and the particle dynamics can not be described in terms
of damping and diffusion coefficients. In the following, we will investigate
the transport properties of a system of non-interacting particles
described by Eqs.\ (\ref{eq:velomed}) and (\ref{eq:gamma}) through
the evaluation of its optical conductivity.

\section{The Optical Conductivity}

The current associated to distinguishable non-interacting particles
described by the equation of motion (\ref{eq:velomed}) reads $j=ev(t)$,
where $j$ satisfy \begin{equation}
\frac{dj}{dt}+\overline{\gamma }\int _{0}^{t}\Gamma (t-t')\, j(t')dt'=\frac{{e^{2}E(t)}}{M}.\label{eq:currenteq}\end{equation}
 This equation may be solved by the well-known method of the Laplace
transform. If we assume that initially there is no current in the
system, Eq.\ (\ref{eq:currenteq}) may be written as $j(z)=\sigma (z)E(z)$,
with \[
\sigma (z)=\frac{{e²}}{M[z+\overline{\gamma }\Gamma (z)]},\]
 where $\Gamma (z)$ is the Laplace transform of $\Gamma (t-t')$.
For all allowed values of $s$, $\Gamma (z)$ can be written as (see
appendix) \begin{eqnarray}
\Gamma (z) & = & \frac{\Omega ^{s+1}F(s,-\frac{\Omega ²}{z²})}{(s+1)z²}\tan \left(\frac{\hbar z}{2kT}\right)\nonumber \\
 &  & -\frac{4k_{B}Tz\Omega ^{s+1}}{\hbar (s+1)}\sum _{n=1}^{\infty }\frac{F(s,-\frac{\Omega ²}{\lambda _{n}²})}{\lambda _{n}²(\lambda _{n}^{2}-z²)},\label{eq:hyper1}
\end{eqnarray}
 where $F$ denotes hyper-geometric functions given by $F(s,x)=\, _{2}F_{1}(1,\frac{1+s}{2},\frac{3+s}{2},x)$
and $\lambda _{n}=(2n-1)\pi k_{B}T/\hbar $, with $n\in \mathbb{N}$.

The optical conductivity may now be promptly obtained by substituting
(\ref{eq:hyper1}) into \begin{equation}
\sigma (\omega )=\lim _{\delta \rightarrow 0^{+}}Re\left[\left.\frac{e^{2}/M}{z+\overline{\gamma }\Gamma (z)}\, \right|_{z=\delta -i\omega }\right].\label{eq:curft}\end{equation}
 Although Eqs.\ (\ref{eq:hyper1})-(\ref{eq:curft}) allow us to
compute $\sigma (\omega )$ for any $s\geq 0$, it may be enlightening
to consider the problem for specific values of $s$, for which the
hyper-geometric series in (\ref{eq:hyper1}) converge to simple functions.
In this way we will be able to proceed analytically in the investigation
of the optical conductivity and obtain, if there are such contributions,
the temperature dependence of the Drude weight and the incoherent
conductivity in the whole frequency range. In the following we will
focus on particular cases illustrating the super-ohmic ($s>1$), ohmic
($s=1$) and sub-ohmic ($s<1$) situations.

\subsubsection{Super-ohmic case, $s=2$}

In this specific case, the hyper-geometric functions involved in Eq.\ (\ref{eq:hyper1})
acquire a simple form \[
_{2}F_{1}(1,3/2,5/2,-x²)=\frac{3}{x}\left(\frac{1}{x}-\frac{\arctan x}{x^{2}}\right),\]
 and the Laplace transform of the damping function reduces to \begin{eqnarray}
\Gamma (z) & = & \frac{4z}{\pi }\sum _{n=1}^{\infty }\frac{(2n-1)\: \arctan \left[\frac{\hbar \Omega }{\pi k_{B}T(2n-1)}\right]}{(2n-1)^{2}-\left(\frac{\hbar z}{\pi k_{B}T}\right)^{2}}\nonumber \\
 &  & -z\, \arctan \left(\frac{\Omega }{z}\right)\tan \left(\frac{\hbar z}{2k_{B}T}\right).\label{eq:gamz1}
\end{eqnarray}
 Substituting then Eq.\ (\ref{eq:gamz1}) into (\ref{eq:curft}),
we obtain the following form for the optical conductivity: \begin{equation}
\sigma (\omega )=\sigma ^{DW}(T)\: \delta (\omega )+\sigma ^{inc}(\omega ,T),\label{eq:cogge}\end{equation}
 where $\sigma ^{DW}(T)$ represents the Drude weight, whereas $\sigma ^{inc}(\omega ,T)$
stands for the incoherent contribution. Let us first analyze $\sigma ^{DW}(T)$,
which is given by \begin{equation}
\sigma ^{DW}(T)=\sigma _{0}\left[1+\frac{4\overline{\gamma }}{\pi }{\displaystyle \sum _{n=1}^{\infty }}\frac{\arctan \left[\frac{\hbar \Omega }{\pi k_{B}T(2n-1)}\right]}{(2n-1)}\right]^{-1}.\label{eq:drudew}\end{equation}
 where $\sigma _{0}=\pi e^{2}/M$ is the Drude weight of the free
particle. Although this sum can not be performed exactly for all temperatures,
we may derive an analytical expression for it in the regime $k_{B}T\gg \hbar \Omega $.
We then find %
\begin{figure}
\begin{center}\includegraphics[  clip,
  scale=0.35]{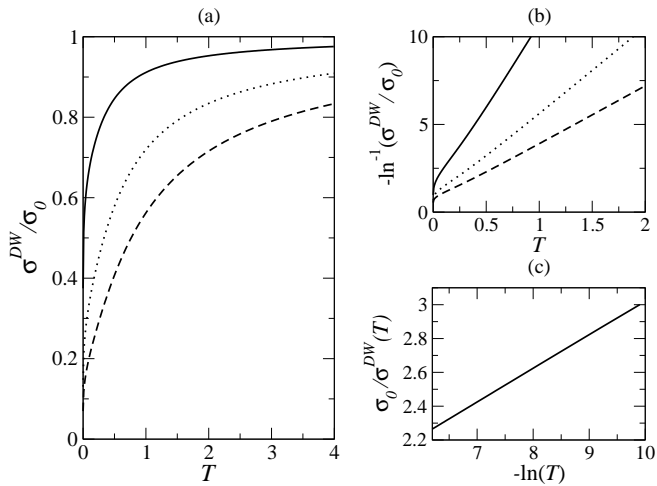}\end{center}

\caption{\label{Drudew}(a) Drude weight as a function of $T$ for $\Omega =1$,
the continuous, dotted and dashed lines correspond to $\overline{\gamma }=0.2$,
$\overline{\gamma }=0.8$ and $\overline{\gamma }=1.6$ respectively.
(b) Temperature behavior of $-1/\ln \sigma ^{DW}$, the continuous,
dotted and dashed lines correspond to $\overline{\gamma }=0.2$, $\overline{\gamma }=0.4$
and $\overline{\gamma }=0.6$ respectively. (c) $1/\sigma ^{DW}$vs
$-\ln T$ for low temperatures and $\overline{\gamma }=0.2$. In all
cases it is assumed $\hbar =k_{B}=1$ and the Drude weight is measured
in units of $\sigma _{0}$.}
\end{figure}

\begin{equation}
\sigma ^{DW}(T)=\frac{{\sigma _{0}}}{1+\frac{4\overline{\gamma }\hbar \Omega }{\pi ²k_{B}T}{\displaystyle \sum _{n=1}^{\infty }}\frac{1}{(2n-1)²}}=\frac{\sigma _{0}}{(1+\overline{\gamma }\frac{\hbar \Omega }{2k_{B}T})},\label{eq:htdrude}\end{equation}
 where the damping term provides a small correction to the particle
mass. Notice that Eq.\ (\ref{eq:htdrude}) correctly reproduces the
free particle behavior at high temperatures. This result is in agreement
with the fact that as the temperature is raised the high frequency
TLSs will play a major role in the composition of the spectral function
which justifies the mass correction as due to the adiabatic approximation.
In other words, the high frequency TLSs adiabatically dress the moving
particle.

A numerical evaluation of the sum involved in Eq. (\ref{eq:drudew})
yields the general behavior of the Drude weight as a function of temperature
(see Fig.\ \ref{Drudew}(a)). Observe that the conductivity grows
with temperature until it saturates at the value of a free particle
in the $T\rightarrow \infty $ limit. In order to investigate the
functional growth of the conductivity with temperature, we have plotted
$-1/\ln \sigma ^{DW}(T)$ vs $T$ in Fig.\ \ref{Drudew}(b). The
plot is linear in $T$, except in the low temperature region. We may
therefore conclude that for high temperatures the Drude weight behaves
as $\sigma ^{DW}\propto \exp [-1/(\alpha +\beta T)]$, where the constants
$\alpha $ and $\beta $ can be determined from the plots and depend
on the values of $\Omega $ and $\overline{\gamma }$. On the other
hand, the vanishing of $\sigma ^{DW}(T)$ as the temperature is lowered
was already expected from (\ref{eq:drudew}), because for strictly
zero temperature the sum to be performed is positive and divergent.

The functional reduction of the conductivity as the temperature decreases
may be determined by considering a certain $\overline{N}=2\overline{n}-1$
such that $\hbar \Omega /\pi k_{B}T\overline{N}\gg 1$. In the limit
of $T\rightarrow 0$ the value of $\overline{N}$ will be of the order
of $\hbar \Omega /k_{B}T$. In this situation Eq. (\ref{eq:drudew})
may be approximately written as \begin{equation}
\sigma ^{DW}(T)\sim \sigma _{0}\left[2\overline{\gamma }\mathcal{S}+\frac{4\overline{\gamma }}{\pi }\sum _{n=\overline{n}}^{\infty }\frac{\arctan \left[\frac{\hbar \Omega }{\pi k_{B}T(2n-1)}\right]}{(2n-1)}\right]^{-1},\label{eq:sigapp}\end{equation}
 where \begin{equation}
\mathcal{S}=\sum _{n=1}^{\overline{n}}\frac{1}{2n-1}=\alpha +\frac{1}{2}\psi ^{(0)}(\overline{n}+1/2),\label{eq:sumadiv}\end{equation}
 with $\alpha =(C+\ln 4)/2$, \[
C=\lim _{m\rightarrow \infty }\sum _{k=1}^{m}\frac{1}{k}-\ln m,\]
 and $\psi ^{(0)}$ denote the polygamma function of zeroth order
given by $\psi ^{(0)}(x)=\partial _{x}\ln \Gamma (x)$. In the zero
temperature limit the sum given by Eq.\ (\ref{eq:sumadiv}) dominates
the behavior of Eq.\ (\ref{eq:sigapp}). Using then the Stirling
expansion and assuming that $\overline{n}$ is of the order of $\hbar \Omega /k_{B}T$,
we finally obtain that the Drude weight in the low temperature limit
behaves as\begin{equation}
\sigma ^{DW}(T\rightarrow 0)\sim \frac{e^{2}}{M\left[\frac{\overline{\gamma }\alpha }{\pi }+\ln \left(\frac{k_{B}T}{\hbar \Omega }\right)^{-\varepsilon \overline{\gamma }}\right]}.\label{eq:appfin}\end{equation}
 In the expression above $\varepsilon $ is a numerical factor that
can determined from the plot of $1/\sigma ^{DW}$vs $-1/\ln T$ for
low temperatures. Inspection of Fig. \ref{Drudew}(c) indicates that
the value of $\varepsilon $ is approximately 1. Therefore we can
conclude that the Drude weight for low temperatures behaves as \begin{equation}
\sigma ^{DW}(T\rightarrow 0)\sim \frac{1}{\ln \left(\frac{\hbar \Omega }{k_{B}T}\right)^{\overline{\gamma }}}.\label{eq:result1}\end{equation}
 Actually the effect of finite $\overline{\gamma }$ can even wash
the Drude weight out as in the na\"{\i}ve Drude model for electric
conductivity of metals. Nevertheless, in our specific model, although
$\sigma ^{DW}$ is reduced, it is still finite even in the presence
of damping for finite $T$. Moreover, as the ratio $\hbar \Omega /k_{B}T$
is directly proportional to the number of TLSs in the lowest energy
state, it is expected that as the temperature is lowered this number
rises, increasing the ability of the particle to lose energy and consequently
leading to a smaller value of $\sigma ^{DW}$.

Now we turn our attention to the behavior of the incoherent part of
the optical conductivity, which is given by \begin{equation}
\sigma ^{inc}(\omega )=\sigma _{0}\frac{\overline{\gamma }\, \tanh \left(\frac{\hbar \omega }{2k_{B}T}\right)\Theta (\Omega -\omega )}{2\omega \left\{ \mathcal{Q}^{2}(\omega ,T)+\left[\frac{\pi \overline{\gamma }}{2}\tanh \left(\frac{\hbar \omega }{2k_{B}T}\right)\right]^{2}\right\} },\label{eq:inchh}\end{equation}
 with \begin{eqnarray}
\mathcal{Q} & = & 1+\frac{4\overline{\gamma }}{\pi }{\displaystyle \sum _{n=1}^{\infty }}\frac{(2n-1)\arctan \left[\frac{\hbar \Omega }{\pi k_{B}T(2n-1)}\right]}{(2n-1)^{2}+(\hbar \omega /\pi k_{B}T)^{2}}\label{eq:pol1}\\
 &  & -\frac{\overline{\gamma }}{2}\tanh \left(\frac{\hbar \omega }{2k_{B}T}\right)\ln \frac{\left|\Omega +\omega \right|}{\left|\Omega -\omega \right|}.\nonumber 
\end{eqnarray}

The presence of the step function in Eq. (\ref{eq:inchh}) ensures
that $\sigma ^{inc}(\omega )$ is exactly zero above the cutoff frequency.
Although for frequencies of the order of $\Omega $ our approach may
provide non accurate results, the zero conductivity in the $\omega >\Omega $
region agrees with the fact that all particle transitions between
states with energy difference larger than $\hbar \Omega $ are forbidden.
This effect is a consequence of having limited the thermal bath phase
space by an abrupt cutoff frequency.

\begin{figure}
\begin{center}\includegraphics[  clip,
  scale=0.35]{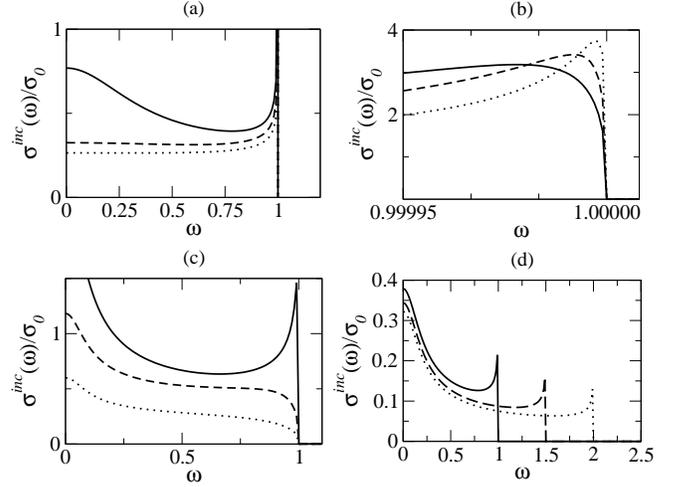}\end{center}

\caption{\label{incs2}(a) $\sigma ^{inc}$ as a function of $\omega $ for
different temperatures with $\overline{\gamma }=0.2$ and $\Omega =1$.
The continuous, dashed and dotted lines correspond to $T=0.09$, $T=0.5$
and $T=2.0$ respectively. (b) Details of $\sigma (\omega )$ near
$\omega =\Omega $. (c) $\sigma ^{inc}$ as a function of $\omega $
for different coupling strengths. The continuous, dashed and dotted
lines correspond to $\overline{\gamma }=0.4$, $\overline{\gamma }=1.5$
and $\overline{\gamma }=3.5$ respectively, with $T=0.02$ and $\Omega =1$.
(d) $\sigma ^{inc}$ vs $\omega $ for $\overline{\gamma }=0.2$ and
$T=0.05$ for different cutoff frequencies, the continuous, dashed
and dotted lines corresponds to $\Omega =1$, $\Omega =1.5$ and $\Omega =2$
respectively. In all cases the optical conductivity is measured in
units of $\sigma _{0}$ and it is assumed $\hbar =k_{B}=1$.}
\end{figure}

In order to discuss the main features of the incoherent part of the
optical conductivity given by Eqs.\
(\ref{eq:inchh})-(\ref{eq:pol1}) we begin by calculating the temperature
dependence of the dc conductivity $\sigma _{dc}=\sigma ^{inc}(\omega =0),$
namely\[
\frac{\sigma _{dc}(T)}{\sigma _{0}}=\frac{\hbar \overline{\gamma }}{4k_{B}T}\left[1+\frac{4\overline{\gamma }}{\pi }{\displaystyle \sum _{n=1}^{\infty }}\frac{\arctan \left[\frac{\hbar \Omega }{\pi k_{B}T(2n-1)}\right]}{(2n-1)}\right]^{-2}.\]
The high temperature limit of this expression can be written as

\[
\frac{\sigma _{dc}(T)}{\sigma _{0}}=\frac{\hbar \overline{\gamma }/4k_{B}T}{\left[1+\frac{\hbar \overline{\gamma }\Omega }{2\pi k_{B}T}\right]^{2}},\]
which correctly goes to zero as the temperature increses in agreement
with the free particle behavior. The interesting point comes from
the divergence at zero temperature. This singular behavior has been
observed in classical non-integrable nonlinear systems where the current
correlation decays to zero in the long time limit, but so slowly,
that the integral over time, which yields the dc conductivity, diverges.\cite{zotos}
In the present case this result can be understood by realizing that
the long time behavior of the damping function (\ref{eq:gamma}) involves
low frequency modes, but the latter do not have enough spectral weight
for the \emph{$s=2$} case to make the current decay.

A further understanding of the problem can be achieved after the analysis
of Fig.\ \ref{incs2}, where the frequency dependence of $\sigma ^{inc}(\omega )$
is plotted for some particular cases. Notice in Fig.\ \ref{incs2}(a)
that as the temperature increases, the conductivity is reduced in
the entire frequency range. This is in agreement with the fact that
as one moves toward the limit in which $k_{B}T\gg \hbar \Omega $
the particle approaches the free behavior and eventually the incoherent
conductivity vanishes, see Eqs.\ (\ref{eq:inchh}) and (\ref{eq:pol1}),
retrieving the free particle result.

Let us now inspect the behavior of $\sigma ^{inc}(\omega )$ when
$\omega \rightarrow \Omega $. Fig. \ref{incs2}(b) shows a zoom of
Fig. \ref{incs2}(a) for $\omega $ around $\Omega $. Near the cutoff
frequency the conductivity reaches a maximum and then continuously
falls to zero as one approaches $\Omega $ from below. This non-monotonic
behavior of the optical conductivity as a function of the frequency
can be basically attributed to memory effects of the damping function
(\ref{eq:gamma}). This effect is responsible for the dephasing of
the different contributions of the bath reaction over the particle.
As a matter of fact, if we reanalyze Eq. (\ref{eq:inchh}), we observe
that the classical impedance $Z(\omega ,T)$ of the reservoir (for
$\omega <\Omega $) corresponds to \[
Z=\sigma _{0}^{-1}\left[\frac{2\omega \mathcal{Q}^{2}(\omega ,T)}{\overline{\gamma }\, \tanh \left(\frac{\hbar \omega }{2k_{B}T}\right)}+\frac{\pi ^{2}\overline{\gamma }\omega }{2}\tanh \left(\frac{\hbar \omega }{2k_{B}T}\right)\right].\]
 Although this expression can not be clearly separated into reactive
and resistive contributions, one of the components of the former will
be \[
\mathcal{X}(\omega ,T)\propto \frac{2\omega \mathcal{Q}^{2}(\omega ,T)}{\overline{\gamma }\, \tanh \left(\frac{\hbar \omega }{2k_{B}T}\right)},\]
 where $\mathcal{Q}(\omega ,T)$ is given by (\ref{eq:pol1}). From
the definition of $\mathcal{Q}(\omega ,T)$, we observe that part
of the reaction of the bath over the particle is nothing but a competition
of terms with a $\pi $ phase difference between them, which causes
this kind of resonance-like behavior. The effect described above does
not appear in the oscillator model, where the bath reacts as a pure
resistance due to its memoryless character, and leads to a monotonic
behavior of $\sigma ^{inc}(\omega )$. It is also expected that as
the coupling strength of the particle-reservoir interaction increases,
the conductivity will decrease. This effect is illustrated in Fig.
\ref{incs2}(c), where it can be seen also that for strong enough
coupling the optical conductivity becomes a monotonic function. It
remains to mention that for a fixed value of the temperature, an increase
in the value of the frequency cutoff leads to a displacement of the
conductivity edge to higher frequencies and to a decrease of $\sigma ^{inc}$
in the whole frequency range. This effect is illustrated in Fig. \ref{incs2}(d)
and it is a consequence of the higher number of states to scatter
the particle, as the value of the cutoff frequency is increased at
a given temperature.

\begin{figure}
\begin{center}\includegraphics[  clip,
  scale=0.30]{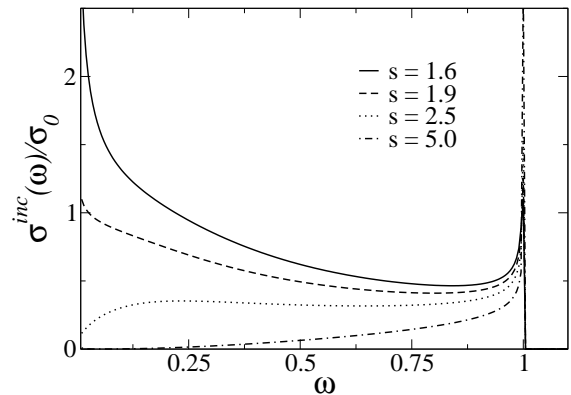}\end{center}

\caption{\label{superO}$\sigma ^{inc}$ as a function of $\omega $ - in
units of $\sigma _{0}$ - for different values of $s$ in the super-ohmic
regime. Notice that as $s$ increases, the value of the $\sigma _{dc}$
decreases for a given frequency. All graphics were generated assuming
$\overline{\gamma }=0.2$, $T=0.09$, $\Omega =1$ and $\hbar =k_{B}=1$. }
\end{figure}

All the previous analysis developed for the $s=2$ specific case illustrates
to some extent the transport properties of the system in the super-ohmic
regime ($s>1$). However, it is worth mentioning that there are significant
differences in the low frequency region of the optical conductivity
as one moves from $1<s<2$ to $s>2$ situation - see Fig.\ \ref{superO}
for instance. The divergence in the dc conductivity is suppressed
for $s>2$ while it persists for $s<2$. This is again a consequence
of the different spectral weights for the low frequency modes as the
value of $s$ is changed.

To conclude this part of the analysis we summarize our main findings.
In the specific case of $s=2$ the system behaves as a perfect conductor
at $T=0$ because of the infinite $\sigma _{dc}$ conductivity. However,
this fact is not reflected in the value of the Drude weight, which
goes to zero as $T$ decreases. This contradiction in the classification
into metallic or insulator states according to the value of the Drude
weight is only apparent. This is so because the current carriers in
our approach are not in an eigenstate of the system invalidating the
characterization of conductors and insulators in terms of the behavior
of $\sigma ^{DW}$. Another point to be mentioned is the fact that
for $s\neq 2$ the dc conductivity, at $T=0$, strongly depends on
the $s$ value, see Fig.\ \ref{superO}. Finally, above the cutoff
$\Omega $, the optical conductivity is always zero at any temperature,
independent of the $s$ value.

\subsubsection{Ohmic case, $s=1$}

This case is of particular importance because it allows us to explicitly
show the differences with the oscillator model discussed in Refs.
{[}\onlinecite{AOphysa}{]} and {[}\onlinecite{AOannal}{]}. In this
specific situation it is useful to notice that \[
_{2}F_{1}(1,1,2,-x²)=\frac{\ln x²+1}{x²}.\]
 Therefore, the general expression (\ref{eq:hyper1}) for the Laplace
transform of the damping function becomes \begin{eqnarray}
\Gamma (z) & = & \frac{1}{2}\tan \left(\frac{\hbar z}{2k_{B}T}\right)\, \ln \left[1+\frac{\Omega ²}{z²}\right]\label{eq:gzsmen1}\\
 &  & -\frac{2\hbar z}{\pi ²k_{B}T}\sum _{n=1}^{\infty }\frac{\ln \left\{ 1+\left[\hbar \Omega /(2n-1)\pi k_{B}T\right]^{2}\right\} }{(2n-1)²-(\hbar z/\pi k_{B}T)²}.\nonumber 
\end{eqnarray}

By substituting (\ref{eq:gzsmen1}) into (\ref{eq:curft}), we obtain
the optical conductivity as a sum of a coherent part given by $\sigma ^{DW}(T)\delta (\omega )$,
where \begin{equation}
\frac{\sigma _{0}}{\sigma ^{DW}}=1+\frac{\hbar \overline{\gamma }}{k_{B}T}\left(\sum _{n=1}^{\infty }\frac{\ln \left\{ 1+\left[\frac{\hbar \Omega }{(2n-1)\pi k_{B}T}\right]^{2}\right\} }{(2n-1)²}+\frac{1}{2}\right),\label{eq:coggem1}\end{equation}
 and an incoherent contribution of the form \begin{equation}
\sigma ^{inc}(\omega )=\frac{(\overline{\gamma }\pi /2)\, \tanh \left(\frac{\hbar \omega }{2k_{B}T}\right)\sigma _{0}\Theta (\Omega -\omega )}{\left[\overline{\gamma }\pi \tanh \left(\frac{\hbar \omega }{2k_{B}T}\right)/2\right]²+\mathcal{R}^{2}(\omega ,T)},\label{eq:incsme1}\end{equation}
 where \begin{eqnarray}
\mathcal{R} & = & \omega +\frac{\overline{\gamma }}{2}\, \tanh \left(\frac{\hbar \omega }{2k_{B}T}\right)\, \ln \left(\frac{\Omega ²-\omega ²}{\omega ²}\right)\label{eq:polsme1}\\
 &  & -\frac{2\overline{\gamma }\hbar \omega }{\pi ²k_{B}T}\sum _{n=1}^{\infty }\frac{\ln \left\{ 1+\left(\hbar \Omega /(2n-1)\pi k_{B}T\right)^{2}\right\} }{(2n-1)²+(\hbar \omega /\pi k_{B}T)^{2}}.\nonumber 
\end{eqnarray}

Once again the Drude weight can not be obtained exactly for all temperatures,
except for $k_{B}T\gg \hbar \Omega $. In this limit we find \begin{equation}
\sigma ^{DW}(T)=\sigma _{0}\left[1+\frac{\pi ²\overline{\gamma }}{96\Omega }\left(\frac{\hbar \Omega }{k_{B}T}\right)^{3}\right]^{-1},\label{eq:htsme1}\end{equation}
 which correctly reproduces the expected free particle conductivity
in the infinite temperature regime, where the incoherent contribution
goes to zero {[}see Eqs.\ (\ref{eq:incsme1}) and (\ref{eq:polsme1}){]}.
If one compares Eqs.\ (\ref{eq:htdrude}) and (\ref{eq:htsme1}),
one realizes that for the same temperature the Drude weight in the
$s=2$ case is lower than in the $s=1$. This is nothing but the effect
of the stronger coupling between the particle and the reservoir modes
as value of $s$ -characterizing the thermal bath- is increased.

The low temperature properties of the Drude weight can be derived
from (\ref{eq:coggem1}) by considering a value of $n=\overline{n}$
such that $\hbar \Omega /k_{B}T\gg \overline{n}$. In this case the
sum involved in (\ref{eq:coggem1}) can be written approximately as
\begin{eqnarray}
\mathcal{S} & = & \ln \left(\frac{\hbar \Omega }{k_{B}T}\right)\sum _{n=1}^{\overline{n}}\frac{1}{(2n-1)^{2}}-\sum _{n=1}^{\overline{n}}\frac{\ln \pi (2n-1)}{(2n-1)^{2}}\nonumber \\
 &  & +\sum _{n=\overline{n}}^{\infty }\frac{\ln \left\{ 1+\left[\frac{\hbar \Omega }{(2n-1)\pi k_{B}T}\right]^{2}\right\} }{(2n-1)²}.\label{eq:sume1}
\end{eqnarray}
 Notice that for $T\rightarrow 0$ the value of $\overline{n}$ goes
to infinity and therefore the first term in the rhs of (\ref{eq:sume1})
dominates the sum. Substituting this result into (\ref{eq:coggem1})
we obtain that for low temperatures the Drude weight goes to zero
as\begin{equation}
\sigma ^{DW}(T\rightarrow 0)\propto \frac{k_{B}T}{\hbar \overline{\gamma }\, \ln \left[\frac{\hbar \Omega }{k_{B}T}\right]}.\label{eq:lowtsme1}\end{equation}
 This expression correctly reproduces the well known result for the
oscillator model,\cite{AOannal} namely there is no coherent contribution
to the conductivity when $\Omega $ goes to infinity. This result
could have been foreseen from the analysis of the particle dynamics
{[}see Eq.\ (\ref{eq:gamma}) for $s=1$ and $T\rightarrow 0${]}\emph{.}
It should be stressed that for $s=1$ $\sigma ^{DW}$ goes to zero
faster than for $s=2$ {[}see Eq.\
(\ref{eq:result1}){]}, as a consequence of the stronger interaction
of the particle with the low energy modes.

\begin{figure}
\begin{center}\includegraphics[  clip,
  scale=0.37]{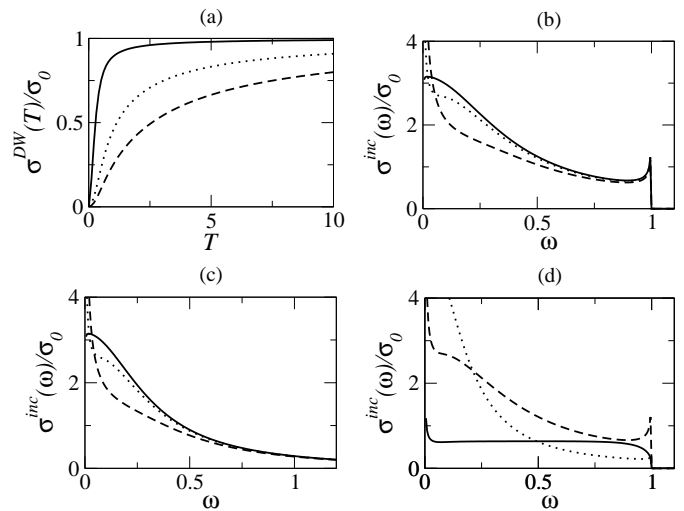}\end{center}

\caption{\label{p-incohvT}(a) $\sigma ^{DW}$ as a function of $T$ for different
coupling strength with $\Omega =1$. The continuous, dotted and dashed
lines correspond to $\overline{\gamma }=0.2$, $\overline{\gamma }=2$
and $\overline{\gamma }=5$ respectively. (b) $\sigma (\omega ,T)$
vs $\omega $ for different temperatures with $\overline{\gamma }=0.2$
and $\Omega =1$. The continuous, dotted and dashed lines correspond
to $T=0.001$, $T=0.02$ and $T=0.09$ respectively. (c) The same
as (b) assuming $\Omega =100$. (d) $\sigma ^{inc}$ vs $\omega $
for $\Omega =1$ and $T=0.02$ for different coupling strength, the
continuous, dashed and dotted lines corresponds to $\overline{\gamma }=1$,
$\overline{\gamma }=0.2$ and $\overline{\gamma }=0.09$ respectively.
In all cases the conductivity is measured in units of $\sigma _{0}$
and it is assumed $\hbar =k_{B}=1$.}
\end{figure}

After having examined the temperature dependence of the Drude weight
for ohmic dissipation in the analytically accessible limits, we proceed
to a numerical evaluation of Eq. (\ref{eq:coggem1}). Fig. \ref{p-incohvT}(a)
shows the behavior of the Drude weight as a function of temperature
for different coupling strength values. Observe that for any given
temperature $\sigma ^{DW}$ decreases as the interaction strength
becomes stronger. Indeed, for stronger interactions the momentum transferred
to the reservoir is larger and consequently the conductivity is reduced.
This also implies that the free particle behavior will be reached
at higher temperatures as $\overline{\gamma }$ increases.

Our next step is the analysis of the incoherent part of the optical
conductivity. The first point to be mentioned is the infinite dc conductivity
of the system at any finite temperature. This result is obtained taking
the limit of zero frequency in the expressions (\ref{eq:incsme1})
and (\ref{eq:polsme1}) and can be explained by the fact that the
dc conductivity is determined by the low frequency modes in the damping
function (\ref{eq:gamma}). We have plotted in Fig.\ \ref{p-incohvT}(b)
$\sigma ^{inc}$ vs $\omega $ for different temperatures with a finite
cutoff frequency $\Omega =1$. The forbidden particle transitions
involving energy exchange above $\hbar \Omega $ leads to zero conductivity
for $\omega >\Omega $, as already obtained in the $s=2$ case. One
may also notice that as the temperature increases the conductivity
grows for small values of $\omega $. This can be understood by recalling
that as the temperature rises, the particle-reservoir interaction
- which can actually change the particle momentum - is less effective
for the low energy modes and consequently the conductivity also rises
for such frequencies. It should be also mentioned that as the temperature
is lowered, $\sigma ^{inc}(\omega )$ acquires a Lorentzian form.
This behavior becomes evident in Fig.\ \ref{p-incohvT}(c), in which
we considered $\Omega =100$. At lower temperatures one approaches
the oscillator model. Indeed, taking the limits $\Omega \rightarrow \infty $
and $T\rightarrow 0$, we obtain\[
\sigma ^{inc}(\omega )=\frac{\pi \overline{\gamma }/2}{\omega ²+(\pi \overline{\gamma }/2)^{2}}\, \sigma _{0}\]
 which is the well known result for the oscillator model. Finally,
Fig. \ref{p-incohvT}(d) shows the effect of varying the coupling
strength. The main feature to be observed is the reduction of the
conductivity as $\overline{\gamma }$ increases, consistent with the
fact that it is then easier for the particle to transfer momentum
to the reservoir.

\subsubsection{Sub-ohmic case, $s=0$}

Another interesting result, which differs considerably from the cases
already discussed, arises when $s=0$. In this case, the hyper-geometric
function reads \[
_{2}F_{1}(1,1/2,3/2,-x²)=\frac{\arctan x}{x}\]
 and the Laplace transform of the damping function given by Eq.\ (\ref{eq:hyper1})
acquires the form \begin{equation}
\Gamma (z)=\frac{4k_{B}Tz}{\hbar }\sum _{n=1}^{\infty }\frac{1}{\lambda _{n}^{2}-z^{2}}\left[\frac{\arctan \left(\frac{\Omega }{z}\right)}{z}-\frac{\arctan \left(\frac{\Omega }{\lambda _{n}}\right)}{\lambda _{n}}\right],\label{eq:ltrax}\end{equation}
 where we have used that $\lambda _{n}=(2n-1)k_{B}T\pi /\hbar $ with
$n$ integer.

Substituting now Eq.\ (\ref{eq:ltrax}) into (\ref{eq:curft}), we
obtain an expression for the optical conductivity which has zero Drude
weight at all temperatures and an incoherent part of the form\begin{equation}
\sigma ^{inc}(\omega )=\sigma _{0}\frac{2\overline{\gamma }\omega \tanh \left(\frac{\hbar \omega }{2k_{B}T}\right)\Theta (\Omega -\omega )}{\mathcal{G}^{2}(\omega ,T)+\left[\pi \overline{\gamma }\tanh \left(\hbar \omega /2k_{B}T\right)\right]^{2}},\label{eq:incxxx}\end{equation}
 with \begin{eqnarray}
\mathcal{G} & = & 2\omega ^{2}-\frac{{8\hbar ^{2}\overline{\gamma }\omega ^{2}}}{\pi ^{3}k^{2}T^{2}}\sum _{n=1}^{\infty }\frac{\arctan \left(\hbar \Omega /N\pi k_{B}T\right)}{N\left[N^{2}+(\hbar \omega /\pi k_{B}T)^{2}\right]}\nonumber \\
 &  & +\overline{\gamma }\, \tanh \left(\frac{\hbar \omega }{2k_{B}T}\right)\: \ln \frac{\left|\Omega -\omega \right|}{\left|\Omega +\omega \right|}.\label{eq:ffx}
\end{eqnarray}

The dc conductivity, obtained by taking the $\omega \rightarrow 0$
limit in Eqs.\ (\ref{eq:incxxx})-(\ref{eq:ffx}), may be written
as $\sigma _{dc}/\sigma _{0}=4k_{B}T/\pi ^{2}\hbar \overline{\gamma }$.
From this result we conclude that the system behaves as an insulator
with zero dc conductivity at $T=0$.

Inspection of Eq.\ (\ref{eq:incxxx}) yields that the incoherent
part of the conductivity is finite only below the cutoff frequency,
just as in the cases discussed previously. Another important feature
is the way in which one may recover the free particle behavior. In
order to analyze this limit it should be noticed that at high temperatures
Eqs. (\ref{eq:incxxx}) and (\ref{eq:ffx}) may be written as \begin{equation}
\sigma ^{inc}(\omega )=\frac{e²}{M}\frac{\zeta }{\omega ^{2}+\zeta ^{2}},\label{eq:sm2ht}\end{equation}
 where $\zeta =\pi \hbar \overline{\gamma }/4k_{B}T$. In the limit
$\zeta \rightarrow 0$, which corresponds to infinite temperature
or zero coupling strength, Eq. (\ref{eq:sm2ht}) goes to $(\pi e²/M)\delta (\omega )$,
correctly reproducing the optical conductivity of a free particle.

\begin{figure}
\begin{center}\includegraphics[  clip,
  scale=0.35]{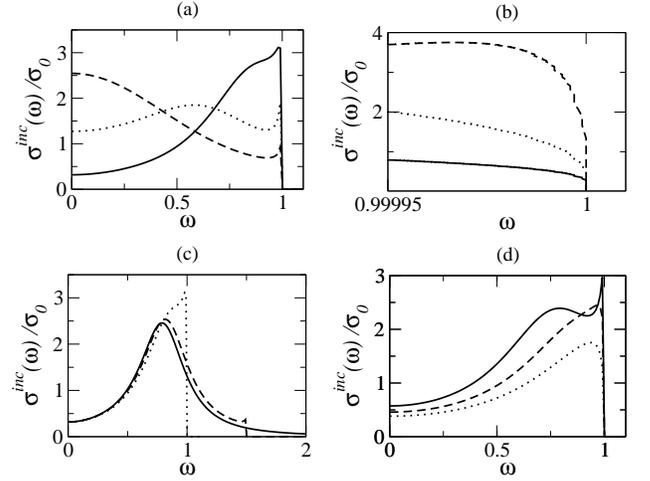}\end{center}

\caption{\label{x-graf1} (a) Incoherent contribution to the optical conductivity
for different temperatures with $\overline{\gamma }=0.2$ and $\Omega =1$.
The continuous, dotted and dashed lines correspond to $T=0.05$, $T=0.2$
and $T=0.4$ respectively. (b) Detail of (a) near the cutoff frequency.
(c) $\sigma ^{inc}$ vs $\omega $ for different values of $\Omega $.
The continuous, dashed and dotted lines corresponds to $\Omega =100$,
$\Omega =1.5$ and $\Omega =1$ respectively, in all cases $\overline{\gamma }=0.2$
and $T=0.05$. (d) $\sigma ^{inc}$ vs $\omega $ for different values
of the coupling strength. The continuous, dashed and dotted lines
corresponds to $\overline{\gamma }=0.2$, $\overline{\gamma }=0.25$
and $\overline{\gamma }=0.3$ respectively, with $\Omega =1$ and
$T=0.09$. In all cases the optical conductivity is measured in units
of $\sigma _{0}$ and it is assumed $\hbar =k_{B}=1$.}
\end{figure}

The general behavior of $\sigma ^{inc}(\omega )$ is displayed in
Fig.\ \ref{x-graf1} for different values of temperature, coupling
strength and cutoff frequency. Fig.\ \ref{x-graf1}(a) shows that
as the temperature increases, one approaches the frequency dependence
given by Eq.\ (\ref{eq:sm2ht}), which has a maximum at $\omega =0$.
In this regime, the classical impedance of the reservoir is purely
resistive, analogous to the oscillator-bath model. However, as the
temperature decreases, the value of $Z(\omega )$ is determined by
a competition between terms completely out of phase, see Eq.\ (\ref{eq:ffx}),
reaching the maximum value for $\omega \neq 0$. For $T\rightarrow 0$,
the frequency at which the conductivity exhibits a maximum becomes
closer to the cutoff frequency, characterizing an extreme non-Drude
behavior. As one approaches $\Omega $ from below, no matter the value
of the temperature, the conductivity smoothly falls to zero. This
effect, which is related to the lack of available states in the reservoir
to scatter the particle, is shown in Fig.\ \ref{x-graf1}(b). It
is also worth mentioning the small variations of the maximum conductivity
as the cutoff frequency is changed, as illustrated in Fig.\ \ref{x-graf1}(c).
Finally, Fig.\ \ref{x-graf1}(d) shows the effect of varying the
value of the coupling strength $\overline{\gamma }$ for a given temperature
and cutoff frequency. An increase in the interaction strength leads
to a reduction of the conductivity. This is an expected result because
for larger values of $\overline{\gamma }$, the particle can transfer
momentum to the reservoir in an effective way, and as a consequence,
the conductivity is reduced.

\begin{figure}
\begin{center}\includegraphics[  clip,
  scale=0.32]{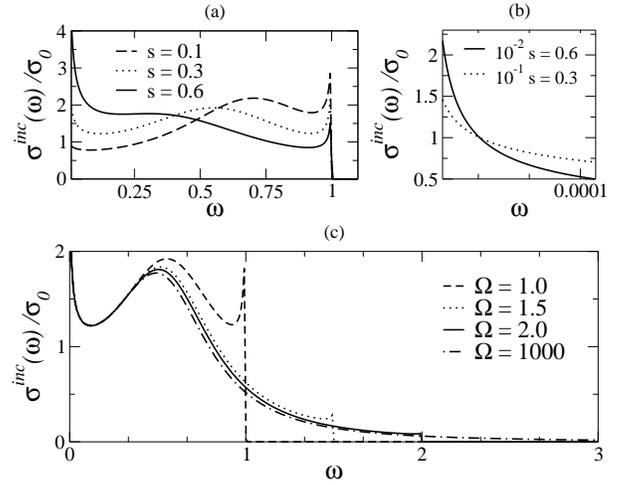}\end{center}

\caption{\label{underdamp} (a) Incoherent contribution to the optical conductivity
in the sub-ohmic regime, in all cases it was assumed $\overline{\gamma }=0.2$,
$T=0.09$ and $\Omega =1$. (b) Detailed form of the conductivity
in (a) near $\omega =0$ for $s=0.6$ and $s=0.3$. (c) $\sigma ^{inc}$
vs $\omega $ for different values of $\Omega $, in all cases $T=0.09$,
$\overline{\gamma }=0.2$ and $s=0.3$. The optical conductivity is
always measured in units of $\sigma _{0}$ and it is assumed $\hbar =k_{B}=1$.}
\end{figure}

At this point we can conclude the $s=0$ analysis pointing out that
in this situation the conductivity is always incoherent and non zero
just below the cut off frequency with a dc value that vanishes as
$T$ goes to zero. In order to verify if those features are common
to the entire sub-ohmic regime, we have numerically evaluated the
combined Eqs.\ (\ref{eq:ltrax}) and (\ref{eq:curft}) for different
values of $s<1$, see Fig.\ \ref{underdamp}(a). As it can be observed,
the general behavior of the conductivity resembles the $s=0$ case
(see Fig.\ 6), except in the low frequency region. This fundamental
difference is illustrated in Fig. \ref{underdamp}(b) showing that
$\sigma ^{inc}$ is singular at $\omega =0$, although our numerical
calculation indicates that the Drude weight remains zero at any finite
temperature and value of $s$. In order to investigate in more detail
the general behavior of $\sigma ^{inc}$ in the low frequency region,
it is useful to notice that near $\omega =0$ the optical conductivity
does not depend on the value of the cut off frequency, see Fig.\
\ref{underdamp}(c). Therefore, we may simply take the limit $\Omega \rightarrow \infty $
in the Laplace transform of the damping function given by expression
(\ref{eq:hyper1}) and obtain an expression for $\Gamma (z)$ for
any $0<s<1$, namely

\begin{equation}
\Gamma (z)=\frac{2\pi k_{B}T\, z}{\hbar \cos (\pi s/2)}\sum _{n=1}^{\infty }\frac{z^{s-1}-\lambda _{n}^{s-1}}{\lambda _{n}^{2}-z^{2}}.\label{eq:undLT}\end{equation}

Substituting the expression above in (\ref{eq:curft}) we obtain the
optical conductivity, which has a zero Drude weight, besides an incoherent
contribution given by

\begin{equation}
\frac{\sigma ^{inc}(\omega )}{\sigma _{0}}=\frac{2\overline{\gamma }\omega ^{s-3}\tanh (\frac{\hbar \omega }{2k_{B}T})}{4\mathcal{K}^{2}(\omega ,s)+\left(\pi \overline{\gamma }\omega ^{s-1}\tanh (\frac{\hbar \omega }{2k_{B}T})\right)^{2}},\label{eq:undsig}\end{equation}
 where\begin{eqnarray}
\mathcal{K} & = & 1+\frac{\pi \overline{\gamma }}{2}\tan (\frac{s\pi }{2})\omega ^{s-2}\tanh \left(\frac{\hbar \omega }{2k_{B}T}\right)\nonumber \\
 &  & -\frac{2\overline{\gamma }(\pi k_{B}T)^{2}}{\hbar ^{2}\cos (\frac{s\pi }{2})}\sum _{n=1}^{\infty }\frac{(2n-1)^{s-1}}{(2n-1)^{2}+(\frac{\hbar \omega }{2k_{B}T})^{2}}.\label{eq:underdefk}
\end{eqnarray}
 This result is in complete agreement with the previous numerical
analysis for finite cutoff frequency and promptly allows us to derive
the way in which the conductivity diverges as one approaches $\omega =0$,

\[
\frac{\sigma ^{inc}(\omega \sim 0)}{\sigma _{0}}=\frac{4k_{B}T\, \cos ^{2}(\pi s/2)}{\pi ^{2}\hbar \overline{\gamma }}\, \omega ^{-s}.\]
 It is worth mentioning that this expression is also valid for the
$s=0$ case (see Eq.\ (\ref{eq:ltrax}), for instance), correctly
reproducing the finite $\sigma _{dc}(T)$ value already found.

With the analysis of the $\Omega \rightarrow \infty $ limit we conclude
that the finite dc conductivity obtained for the $s=0$ case is an
exception within the sub-ohmic regime, in which the dc conductivity
is, in general, divergent. This discussion summarizes our main findings
in the study of the optical conductivity for cases in which $0\leq s<1$.

\section{Conclusions}

We have studied the transport properties of non-interacting particles
coupled to a set of two-level systems described by a temperature-dependent
spectral function. Our approach was based on the Feynman-Vernon functional-integral
formalism which allows us to trace out the two-level system coordinates
and obtain an effective equation of motion for the particle in terms
of the phenomenological TLSs spectral function.

The evaluation of the transport properties were performed considering
different values of the power $s$ of the generic spectral function,
which is proportional to $\omega ^{s}$. Two different regimes were
found; the super-ohmic ($s>1$) and ohmic ($s=1$) in which the optical
conductivity has both, coherent and incoherent contributions and the
sub-ohmic ($s<1$) where the conductivity is strictly incoherent.
In the first case the Drude weight goes to zero as the temperature
decreases in a logarithmic fashion and the incoherent part shows a
strong non-Drude behavior with a divergent dc conductivity for $T=0$,
except for the ohmic case in which the zero temperature dc conductivity
is finite. In the entire sub-ohmic regime, the Drude weight is always
zero at all temperatures. In contrast, the dc conductivity is only
finite in the $s=0$ specific case, going linearly to zero as the
temperature decreases, whereas in general it is divergent at all temperatures.
It should be mentioned that in this regime, the general behavior of
$\sigma (\omega )$ is strongly non-Drude, with the highest conductivity
arising at finite values of $\omega $. All these properties strongly
differ from the simple behavior induced by an oscillator thermal bath
on a system of non-interacting particles. The latter exhibits no temperature
dependence and zero Drude weight with a finite dc conductivity. All
this results are summarized in the table above\emph{.} Concerning
the limitation of the approach employed here, it is important to stress
that it is accurate for long times (or low frequencies) compared with
the response time of the system, which is basically $\Omega ^{-1}$.
Therefore in all cases where the cutoff frequency was assumed to be
finite, the values of the optical conductivity near $\omega =\Omega $
should be interpreted carefully.

Finally, it is worth pointing out that the model analyzed here could
be relevant for the study of the dynamics of a particle in the presence
of a distribution of locally quartic plus quadratic potentials. We
expect that this simple model may contribute to the understanding
of the transport properties of low dimensional systems at low temperatures,
where the physics is often dominated by defects and impurities.

\begin{widetext}

\begin{figure}
\begin{center}\begin{center}

\begin{tabular}{|c|c|c|c|c|c|}
\hline \hspace{3.2cm}
&
\(\, \sigma _{dc}\, \)&
\(\; \sigma _{T\rightarrow 0}^{DW}\; \)&
\(\; \sigma _{T\rightarrow \infty }^{DW}\; \)&
\(\; \sigma _{T\rightarrow 0}^{inc}\; \)&
\(\; \sigma _{T\rightarrow \infty }^{inc}\; \)\\
\hline
Super-Ohmic&
diverges for \(T\rightarrow 0\footnote{For $s\leq 2$, in other cases it is finite}\)&
zero&
\(\sigma _{0}\)&
Anomalous insulator - non Drude&
zero\\
Ohmic&
diverges for finite \(T\)\footnote{For $\Omega \rightarrow \infty$ reproduces the oscillator results}

&
zero&
\(\sigma _{0}\)&
Insulator - Lorentzian form$^b$&
zero\\
Sub-Ohmic&
diverges for finite $T$\footnote{Except in the s=0 case, where the \(\sigma_{dc}\) goes to zero}&
-&
-&
Perfect insulator - non Drude&
\(\sigma _{0}\delta (\omega )\)\\
\hline

\end{tabular}

\end{center}\end{center}
\end{figure}

\end{widetext}

\section{Acknowledgments}

AVF would kindly acknowledge the financial support from Funda\c{c}\~{a}o
de Amparo \`{a} Pesquisa do Estado de S\~{a}o Paulo (FAPESP) during
his post-doctoral period at the Universidade Estadual de Campinas
and AOC the partial support from Conselho Nacional de Desenvolvimento
Cient\'{\i}fico e Tecnol\'{o}gico (CNPq). We are grateful to the
Swiss National Foundation for Scientific Research, grant 620-62868.00,
which allowed for the establishment of this collaboration.

\section*{Appendix \label{laplace}}

In order to calculate the Laplace transform of $\Gamma (t)$ we will
use the expansion\cite{Gradshtein} \[
\tanh \left(\frac{\pi x}{2}\right)=\frac{4x}{\pi }{\displaystyle \sum _{n=1}^{\infty }\frac{1}{(2n-1)²+x²}}.\]
 which allows us to rewrite $\Gamma (z)$ as \begin{equation}
\Gamma (z)=\frac{4k_{B}Tz}{\hbar }\sum _{n=0}^{\infty }\mathbb{I}(n,\Omega ),\label{eq:gamdef}\end{equation}
 where \[
\mathbb{I}(n,\Omega )=\int _{0}^{\Omega }\frac{\omega '^{s}\, d\omega '}{(\omega '^{2}+z²)(\omega '²+\lambda _{n}^{2})},\]
 with $\lambda _{n}=(2n-1)\pi k_{B}T/\hbar $ and $n\in \mathbb{N}$.
This integral can be split into two terms, \[
\mathbb{I}=\frac{1}{z²-\lambda _{n}^{2}}\left[z²\int _{0}^{\Omega }\frac{\omega '^{s-2}\, d\omega '}{(\omega '^{2}+z²)}-\lambda _{n}^{2}\int _{0}^{\Omega }\frac{\omega '^{s-2}\, d\omega '}{(\omega '^{2}+\lambda _{n}^{2})}\right].\]
 The new integrals are easily performed, yielding \[
\int _{0}^{\Omega }\frac{\omega '^{s-2}\, d\omega '}{(\omega '^{2}+\nu ^{2})}=\frac{\Omega ^{s-1}}{\nu ²}\sum _{m=0}^{\infty }\frac{(-1)^{m}}{2m+s-1}\left(\frac{\Omega }{\nu }\right)^{2m}.\]
 We now rewrite the expression for $\mathbb{I}(n,\Omega )$ as \begin{eqnarray}
\mathbb{I}(n,\Omega ) & = & \frac{\Omega ^{s+1}}{(\lambda _{n}^{2}-z²)}\left[\frac{1}{z²}\sum _{m=1}^{\infty }\frac{(-1)^{m-1}}{2m+s-1}\left(\frac{\Omega }{z}\right)^{2m-2}\right.\nonumber \\
 &  & \left.-\frac{1}{\lambda _{n}^{2}}\sum _{m=1}^{\infty }\frac{(-1)^{m-1}}{2m+s-1}\left(\frac{\Omega }{\lambda _{n}}\right)^{2m-2}\right].\label{eq:serie1}
\end{eqnarray}
 Substituting this equation into (\ref{eq:gamdef}) and expressing
the sums involved in (\ref{eq:serie1}) in terms of the hyper-geometric
functions we obtain \begin{eqnarray}
\Gamma (z) & = & \frac{4k_{B}Tz}{\hbar (s+1)}\sum _{n=1}^{\infty }\frac{\Omega ^{s+1}}{\lambda _{n}^{2}-z²}\left[\frac{_{2}F_{1}(1,\frac{1+s}{2},\frac{3+s}{2},-\frac{\Omega ²}{z²})}{z²}\right.\nonumber \\
 &  & \left.-\frac{_{2}F_{1}(1,\frac{1+s}{2},\frac{3+s}{2},-\frac{\Omega ²}{\lambda _{n}²})}{\lambda _{n}²}\right].\label{eq:hyper2}
\end{eqnarray}
 By performing the sum in the first term of Eq.\ (\ref{eq:hyper2})
and introducing the notation $_{2}F_{1}(1,\frac{1+s}{2},\frac{3+s}{2},x)=F(s,x)$,
we finally obtain Eq.\ (\ref{eq:hyper1}).

\end{document}